\documentclass{optica-article}
\journal{opticajournal}
\articletype{Research Article}
\begin{document}

%%% Front Matter %%%
\title{Reconfigurable quantum photonic circuits based on quantum dots}

\author{Adam McCaw,\authormark{1}, Jacob Ewaniuk\authormark{1}, Bhavin J. Shastri,\authormark{1,2}, and Nir Rotenberg\authormark{1,*}}

\address{\authormark{1}Centre for Nanophotonics, Department of Physics, Engineering Physics, \& Astronomy, Queen's University, 64 Bader Lane, Kingston, Ontario, Canada K7L 3N6}
\address{\authormark{2}Vector Institute, Toronto, Ontario, Canada M5G 1M1}

\email{\authormark{*}nir.rotenberg@queensu.ca}

\begin{abstract*}
Quantum photonic integrated circuits, composed of linear-optical elements, offer an efficient way for encoding and processing quantum information on-chip. At their core, these circuits rely on reconfigurable phase shifters, typically constructed from classical components such as thermo- or electro-optical materials, while quantum solid-state emitters such as quantum dots are limited to acting as single-photon sources. Here, we demonstrate the potential of quantum dots as reconfigurable phase shifters. We use numerical models based on established literature parameters to show that circuits utilizing these emitters enable high-fidelity operation and are scalable. Despite the inherent imperfections associated with quantum dots, such as imperfect coupling, dephasing, or spectral diffusion, our optimization shows that these do not significantly impact the unitary infidelity. Specifically, they do not increase the infidelity by more than 0.001 in circuits with up to 10 modes, compared to those affected only by standard nanophotonic losses and routing errors. For example, we achieve fidelities of 0.9998 in quantum-dot-based circuits enacting controlled-phase and -not gates without any redundancies. These findings demonstrate the feasibility of quantum emitter-driven quantum information processing and pave the way for cryogenically-compatible, fast, and low-loss reconfigurable quantum photonic circuits. 
\end{abstract*}

%%% Body %%%
\section{Introduction}
Reconfigurable quantum photonic integrated circuits (qPICs) are versatile tools capable of simulating molecular dynamics \cite{sparrow2018simulating_molecular_ref}, executing quantum logic \cite{carolan2015universal}, and generating multidimensional entanglement \cite{wang2018multidimensional_entanglement_1, adcock2019programmable_entanglement_2, llewellyn2020chip_entanglement_3, vigliar2021error_entanglement_4, bao2023very_2500_entanglement_5}. They utilize quantum properties such as entanglement and indistinguishability for information processing, which is unachievable through classical means. This capability is crucial for developing emerging quantum communication and computation technologies.

\begin{figure}[ht!]
\centering\includegraphics[scale=0.35]{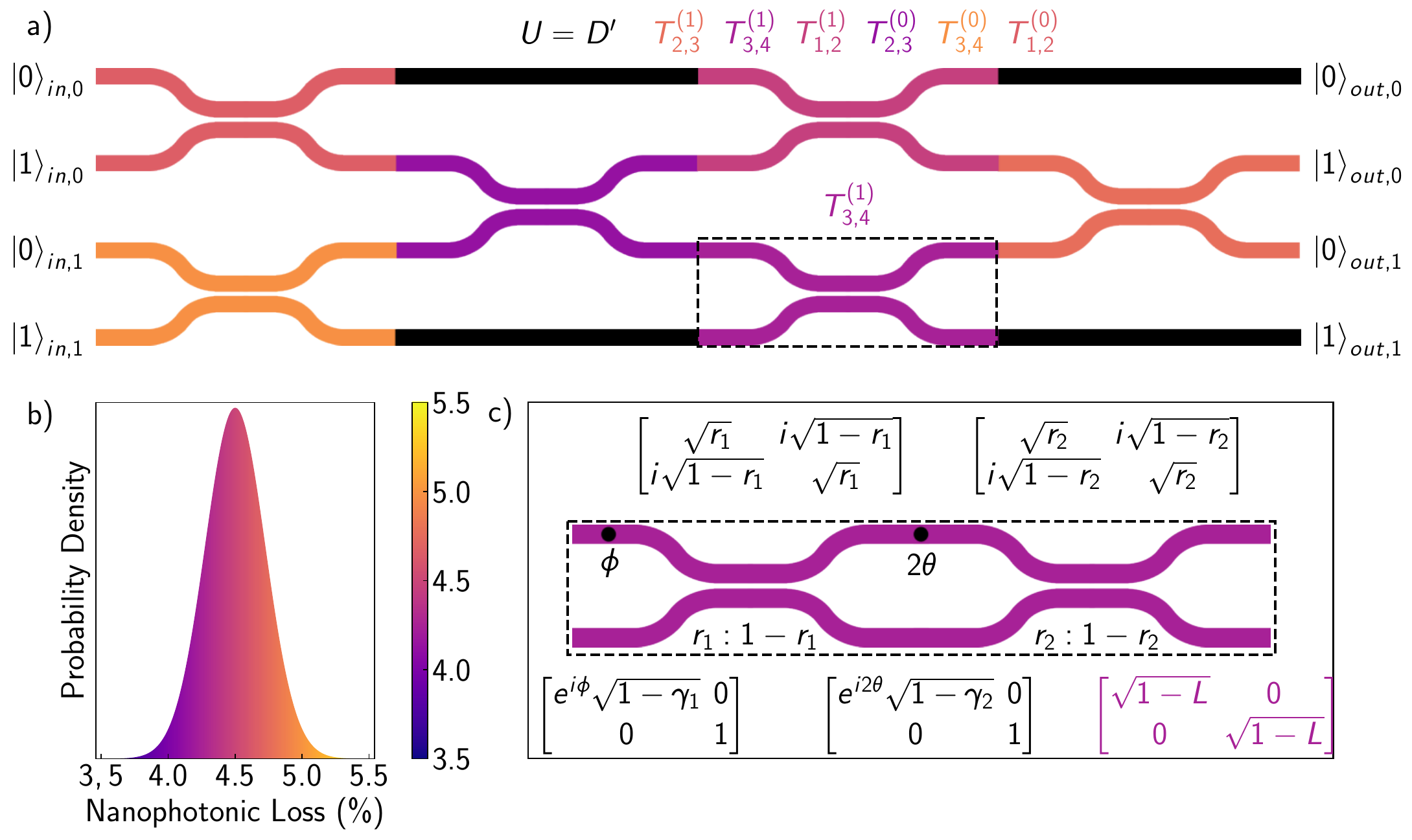}
\caption{\label{fig:schematic}Scalable Reconfigurable MZI Meshes. (a) 4-mode MZI mesh colored to represent nanophotonic losses in each interferometer. (b) Normal distribution for nanophotonic loss with a mean of $4.5\%$ and standard deviation that is $5\%$ of the mean. (c) A zoom-in on the basic component of the mesh (dashed region in a), a MZI with two phase shifters $\phi$ and $2\theta$ and two imperfect beam splitters along with the imperfect transfer matrices for each component and for the overall nanophotonic loss.}
\end{figure}

To date, qPICs have predominantly operated at room temperature, harnessing the advancements of foundry photonics to create increasingly complex devices \cite{bogaerts2020programmable_QPIC_Rev_1, hou2023hardware_QPIC_Hardware_Correction}. As depicted in Fig.~\ref{fig:schematic}, the core of these circuits is a mesh of Mach-Zehnder interferometers (MZIs) \cite{clements2016optimal}, where each MZI is comprised of two directional couplers and two phase shifters (see Fig.~\ref{fig:schematic}b), typically thermo-optical in nature \cite{parra2020ultra_low_loss_thermo_optic_ps, jacques2019optimization_fast_thermo_optic_ps}. 
That is, this mesh is fully classical, based on low-loss and highly accurate (albeit slow and hot) phase shifters. Though it is susceptible to unbalanced nanophotonic imperfections, such as some losses and imperfect routing, high-fidelity operation is achievable with redundancy and optimization \cite{Miller:Self_Config, Mower_Englund_CNOT_Paper, Wilkes:16, Pai_Matrix_Opt, Bandyopadhyay:21, shokraneh_theoretical_2021}.

The quantum nature of qPICs stems from the individual photons that propagate through the circuits. Single photons are often generated by optical nonlinearities such as spontaneous parametric down-conversion, which is compatible with room-temperature chips but inherently probabilistic \cite{wagenknecht2010experimental_SPDC, kaneda2016heralded_SPDC_2, couteau2018spontaneous_SPDC_Review}. In contrast, single-photon sources based on solid-state quantum emitters, such as the quantum dots (QDs) we consider, can operate on-demand but require cryogenic temperatures \cite{somaschi2016near_single_photon_QD_1, senellart2017high_single_photon_QD_2, daveau2017efficient_single_photon_QD_3, uppu2020scalable_single_photon_QD_4}. Similarly, high-efficiency integrated single-photon detectors also necessitate a cryogenic environment \cite{pernice2012high_SPD_1, rosenberg2013high_SPD_2, esmaeil2017single_SPD_3}. As a result, both state-of-the-art sources and detectors cannot be heterogeneously integrated with qPICs and instead currently rely on lossy interconnects \cite{wang2023deterministic_Pan}.

Here, we propose that QDs can be used not only as single-photon sources, but also as reconfigurable phase shifters for creating fast, cryogenically-compatible meshes. Our model of QD-based qPICs incorporates standard nanophotonic imperfections, such as losses and routing errors, along with QD-specific non-idealities, like imperfect interactions and both fast and slow noise processes. We use this model to evaluate the fidelity of both the resultant unitary operations and the desired output states. Our findings reveal that these QD-based meshes can be optimized to achieve remarkable scalability, with a unitary infidelity less than $0.001$ for circuits up to $10\times10$ in dimension, using state-of-the-art QD parameters from the literature. We further consider QD-based controlled-phase and -not gates as examples, where we find that state-of-the-art circuits process logical states with fidelities of 0.9998. In sum, our results offer a roadmap to cryogenically-compatible, reconfigurable qPICs based on solid-state quantum emitters.

\section{Quantum-emitter phase shifters}
\begin{figure}[ht!]
\centering\includegraphics[width=\textwidth]{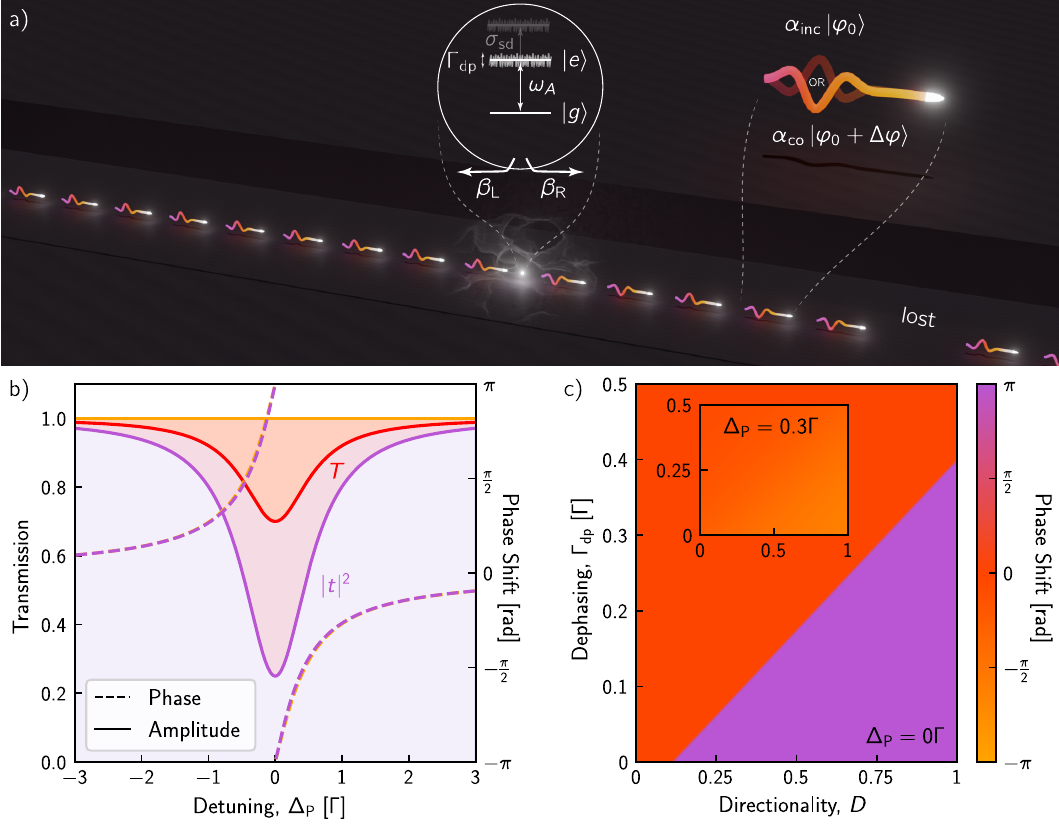}
\caption{\label{fig:QDshifter} Waveguide-coupled chiral quantum dots (QDs). (a) A QD chirally-coupled to a waveguide with its forward direction to the right, $\beta_{\mathrm{R}} \gg \beta_{\mathrm{L}}$, imparts a phase shift $\Delta\varphi$ on coherently-scattered photons but not to those with which it interacts incoherently (due to fast dephasing). If $\beta_{\mathrm{R}}< 1$, then some photons are either reflected or scattered out of the waveguide. (b) Transmission spectrum of the interaction shown in (a). In the ideal case ($\beta_{\mathrm{R}} =1$ and no noise, solid orange curve) all photons are transmitted, while if $\beta_{\mathrm{R}} =0.9$ and $\Gamma_{\mathrm{dp}}=0.1\Gamma$ then the presence of the QD will be imprinted on the total transmission spectrum (solid red curve) and the coherent transmission spectrum (solid purple curve). The associated phase shift applied to the coherently-scattered photons in both the ideal (orange) and non-ideal (purple) scenarios is shown in dashed curve (right axis), which are nearly identical. (c) On-resonance phase shift as a function of directionality and dephasing rate (inset: similar map for a photon-emitter detuning of $\Delta_{\mathrm{p}} = 0.3\Gamma$). The large region over which $\Delta\varphi = \pi$ demonstrates the robustness of QD-based phase shifters.}
\end{figure}

The scattering of photons from a quantum emitter embedded in a nanophotonic waveguide is a complex process \cite{sheremet2023waveguide_Iorsh_QED_Review} that may modulate the photons' amplitude or phase \cite{le2021experimental_Hanna_PRL, staunstrup2023direct} or, when more than a single photon is present, induce complex correlations \cite{shen2007strongly_theory_TLS_Two_Photon,le2022dynamical_Hanna_Nature_Physics}. The exact response depends on the properties of the emitter and the efficiency with which it couples to the various available modes, as sketched in Fig.~\ref{fig:QDshifter}a, yet in the most general case, an input single-photon state with phase $\varphi_0$, $\left|1\right\rangle _{\mathrm{in}}=\left|\varphi_{0}\right\rangle$, will scatter into the mixed state,
\begin{equation}\label{eq:1out}
    \left|1\right\rangle _{\mathrm{out}}=\left\{ \begin{array}{l}
\alpha_{\mathrm{co}}\left|\varphi_{0}+\Delta\varphi\right\rangle \\
\alpha_{\mathrm{inc}}\left|\varphi_{0}\right\rangle 
\end{array}\right. ,
\end{equation}
where, in the presence of losses the sum of the probabilities to scatter coherently and incoherently, $|\alpha_{\mathrm{co}}|^{2}$ and $|\alpha_{\mathrm{inc}}|^{2}$, do not sum to unity.

The description of Eq.~\ref{eq:1out} only holds if the single-photon pulse is not reshaped during scattering, requiring that the pulse linewidth $\sigma_{\mathrm{p}}$ be much shorter than that of the emitter. More formally, we require that $\sigma_{\mathrm{p}} \leq \Gamma / 1000$, ensuring the phase shifter response is linear (see SI for more information).

In this regime, the single-photon transmission coefficient $t$ and total transmission $T$ are the same as that of a weak coherent beam (see SI for derivation),
\begin{eqnarray}
    t&=&1-\Gamma\beta_{\mathrm{R}}\frac{\Gamma_{2}+i\Delta_{\mathrm{p}}}{\Gamma_{2}^{2}+\Delta_{\mathrm{p}}^{2}},\label{eq:t} \\
    T&=&1-2\Gamma\Gamma_{2}\frac{\beta_{\mathrm{R}}(1-\beta_{\mathrm{R}})}{(\Gamma_{2}^{2}+\Delta_{\mathrm{p}}^{2})},\label{eq:T}
\end{eqnarray}
where $\beta_{\mathrm{R}}$ is the coupling efficiency for right-traveling photons (with a total coupling efficiency $\beta = \beta_{\mathrm{R}} + \beta_{\mathrm{L}}$), $\Delta_{\mathrm{p}}$ is the detuning between the photon and emitter-transition frequencies, $\Gamma$ is the decay rate of the emitter, and $\Gamma_{2}=\Gamma / 2 + \Gamma_{\mathrm{dp}}$ where $\Gamma_{\mathrm{dp}}$ is the pure dephasing rate (i.e. fast noise). We note that in the presence of dephasing, $T \neq \left|t\right|^2$ (c.f. Fig.~\ref{fig:QDshifter}b), and that the emitter might also suffer from slow noise leading to spectral diffusion with a characteristic linewidth $\sigma_{\mathrm{sd}}$.

Eqs.~\ref{eq:t} and~\ref{eq:T} allow us to quantify the results of the scattering. The coefficients, $\alpha_{\mathrm{co}}$ and $\alpha_{\mathrm{inc}}$, are related to the transmission, as shown in Fig.~\ref{fig:QDshifter}b. In the ideal case, where $\beta_{\mathrm{R}}=1$ and $\Gamma_{\mathrm{dp}}=0$, the transmission is always unity (orange curve) meaning that $\alpha_{\mathrm{co}}=1$ and $\alpha_{\mathrm{inc}}=0$. Conversely, in the presence of losses and/or dephasing, the situation is more complex with $\alpha_{\mathrm{co}}= \left|t\right|^2$ and $\alpha_{\mathrm{inc}}= T - \left|t\right|^2$ as shown by the purple and red curves.

The phase of the coherently-scattered photons is likewise calculated from $\Delta \varphi = \mathrm{arg}\left(t\right)$, here shown in dashed curves in Fig.~\ref{fig:QDshifter}b corresponding to the ideal and non-ideal scenarios. As can be seen, the imparted phase shift is nearly identical in both cases, spanning the full $2\pi$ and demonstrating the robustness of emitter-based phase shifters. Full, 2D maps of the induced phase shift on resonance, $\Delta_{\mathrm{p}} = 0$, and at $\Delta_{\mathrm{p}}= 0.3\Gamma$, are shown in Figs.~\ref{fig:QDshifter}c and d, respectively, again demonstrating that a $2\pi$ phase change is possible even when the directionality is below $D < 0.25$ or the dephasing rate is above $\Gamma_{\mathrm{dp}} > 0.3\Gamma$. However, this range also depends on emitter parameters such as $\beta$. Corresponding maps of $\left|\alpha_{\mathrm{co}}\right|^2$ are presented in the SI. Together, these enable us to pick an emitter detuning for each desired phase shift, and then calculate the scattered state (c.f. Eq.~\ref{eq:1out}).

\section{QD-based qPICs}
\begin{figure}[ht!]
\centering\includegraphics[width = \textwidth]{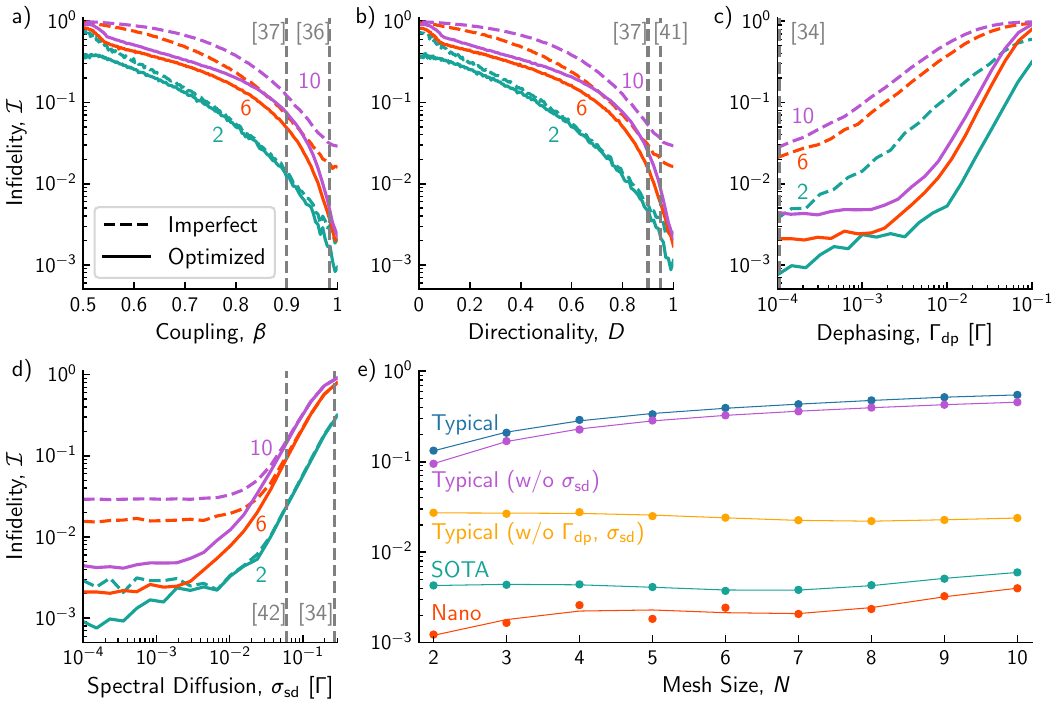}
\caption{\label{fig:scaling}The effects of QD imperfections on circuit unitary infidelity. All plots have 4.5\% nanophotonic loss and 4\% beam splitter error. (a-d) With no other QD imperfections, we examine the effects of (a) coupling efficiency, $\beta$, (b) directionality, $D$, (c) pure dephasing rate, $\Gamma_{\mathrm{dp}}$, and (d) spectral diffusion, $\sigma_{\mathrm{sd}}$  on infidelity, showing both the imperfect (dashed) and optimized (solid) results. In all cases, state-of-the-art QD parameters from literature are indicated by vertical dashed lines. (e) The infidelity of QD-based qPICs as a function of mesh size ranging from $N = 2$ to $10$ for circuits suffering only from nanophotonic imperfections (red), as well as those that additionally suffer from state-of-the-art QD imperfections (green) or typical QD imperfections (blue). See Table~\ref{Tab:QD_Params} for QD parameter values. Also shown are infidelities for circuits based on typical QDs but no noise (orange) or with only fast dephasing (purple), demonstrating that circuit errors are largely due to dephasing.}
\end{figure}

Having seen that quantum emitters such as QDs can serve as reconfigurable phase shifters, we quantify the performance of qPICs based on this technology. To do so, we first compare how well we can reproduce any unitary (i.e. operation) with our emitter-based qPICs relative to the ideal, summarizing the results in Fig.~\ref{fig:scaling}. Here, we show the dependence of the mean circuit infidelity $\mathcal{I}$ (i.e. error) as a function of (a) $\beta$, (b) $D$, (c) $\Gamma_{\mathrm{dp}}$ and (d) $\sigma_\mathrm{sd}$, where, in all, $\mathcal{I}$ is limited by the nanophotonic errors as noted in the caption.

As an example, consider the $\beta$-dependence of $\mathcal{I}$, shown in Fig.~\ref{fig:scaling}a. In dashed curves, we show the infidelity of the non-ideal unitary $\mathrm{U}_{\mathrm{non}}$, constructed following the procedure of \cite{clements2016optimal} using the set of phases $\left\{ 2\theta_{i},\phi_{i}\right\}$ (c.f. Fig.~\ref{fig:schematic}c) calculated for the ideal circuit, but with imperfections subsequently added (see SI for details). This corresponds to the offline training of a photonic circuit. For each coupling efficiency $\beta$, we compare 100 non-ideal unitaries $\mathrm{U}_\mathrm{non}$ to the ideal $\mathrm{U}$ to calculate \cite{clements2016optimal}
\begin{equation}\label{eq:Inf}
\begin{split}
\mathcal{I} &= 1 - \left|\frac{\mathrm{tr}(\mathrm{U}^{\dagger}\mathrm{U}_{\mathrm{non}})}{\sqrt{N \mathrm{tr}(\mathrm{U}_{\mathrm{non}}^{\dagger}\mathrm{U}_{\mathrm{non}})}}\right|^{2} 
\end{split} ,
\end{equation}
which we then average. In the figure, we show the cases for $N = 2,6,10$ mode circuits, where in all cases, we observe a monotonic increase from a baseline $\mathcal{I}$ due to the nanophotonic imperfections when $\beta = 1$, to near-unity error when the coupling efficiency is $\beta = 0.5$. 

Encouragingly, we can optimize the performance of the emitter-based qPICs, following a fast and efficient routine that finds the optimal set $\left\{ 2\theta_{i},\phi_{i}\right\}$ at once \cite{powell2009bobyqa}. This procedure results in an optimized unitary $\mathrm{U}_{\mathrm{opt}}$ that, together with Eq.~\ref{eq:Inf}, enables us to again calculate $\mathcal{I}$, which we show as solid curves in Fig.~\ref{fig:scaling}. In all cases, we observe a significant reduction in errors due to the optimization, with this reduction particularly pronounced for larger circuit sizes and in near-optimal conditions. For example, in Fig.~\ref{fig:scaling}a (and b-d), as $\beta \rightarrow 1$  corresponding to state-of-the-art performance (dashed line) \cite{coupling_98}, we observe almost no increased infidelity as the circuit size increases. A similar dependence is observed as $D$ and $\Gamma_{\mathrm{dp}}$ are scanned (Fig.~\ref{fig:scaling}b and c, respectively), while spectral diffusion only begins affecting the performance when $\sigma_\mathrm{sd} \gtrsim 0.01 \Gamma$ (Fig.~\ref{fig:scaling}d).

Overall, we summarize the circuit scaling in Fig.~\ref{fig:scaling}e, where we plot the raw and optimized infidelity as a function of circuit size, both using typical and state-of-the-art parameters (see Table \ref{Tab:QD_Params}). For typical values (blue curve), we see that the infidelity quickly approaches unity, yet by adding imperfections sequentially (purple and orange curves), we see that this is almost entirely caused by the residual dephasing. This is consistent with the optimized state-of-the-art qPIC performance (green curve), where $\mathcal{I} < 0.006$ is observed for all circuits simulated (up to $N = 10$), as several recent experiments based on QDs have measured $\Gamma_{\mathrm{dp}} \approx 0$ \cite{le2022dynamical_Hanna_Nature_Physics,sollner_chiral_QD_2015,langbein2004radiatively_Low_dephasing,matthiesen2012subnatural_0_Dephasing,nguyen2011ultra_0_dephasing_2}.

\begin{table}[htbp]
\centering
\caption{\label{Tab:QD_Params}State-of-the-art and typical quantum dot parameters. Listed parameters include coupling, $\beta$, directionality, $D$, dephasing, $\Gamma_{\mathrm{dp}}$, spectral diffusion detuning standard deviation, $\sigma_\mathrm{sd}$. For more details on these parameters, see SI.}
\begin{tabular}{ccccc}
\hline
\enspace & Typical & Ref(s) & State-of-the-art & Ref(s) \\
\hline
$\beta$ & 0.90 & \cite{sollner_chiral_QD_2015} & 0.98 & \cite{coupling_98}\\
$D$ & 0.90 & \cite{sollner_chiral_QD_2015} & 0.95 & \cite{coles_chirality_nanobeam_wg}  \\
$\Gamma_{\mathrm{dp}}$ & $0.01\Gamma$ & \cite{le2022dynamical_Hanna_Nature_Physics} & $0\Gamma$ & \cite{le2022dynamical_Hanna_Nature_Physics,sollner_chiral_QD_2015,langbein2004radiatively_Low_dephasing,matthiesen2012subnatural_0_Dephasing,nguyen2011ultra_0_dephasing_2} \\
$\sigma_\mathrm{sd}$ & $0.06\Gamma$ & \cite{thyrrestrup_Spectral_Diffusion_6} & $0\Gamma$ &  \cite{pedersen_pin_QD_control, kuhlmann2013charge_fast_noise_SD} \\
\hline
\end{tabular}
  \label{tab:shape-functions}
\end{table}

\section{Examples: CZ gate}
\begin{figure}[ht!]
\centering\includegraphics[scale=0.8]{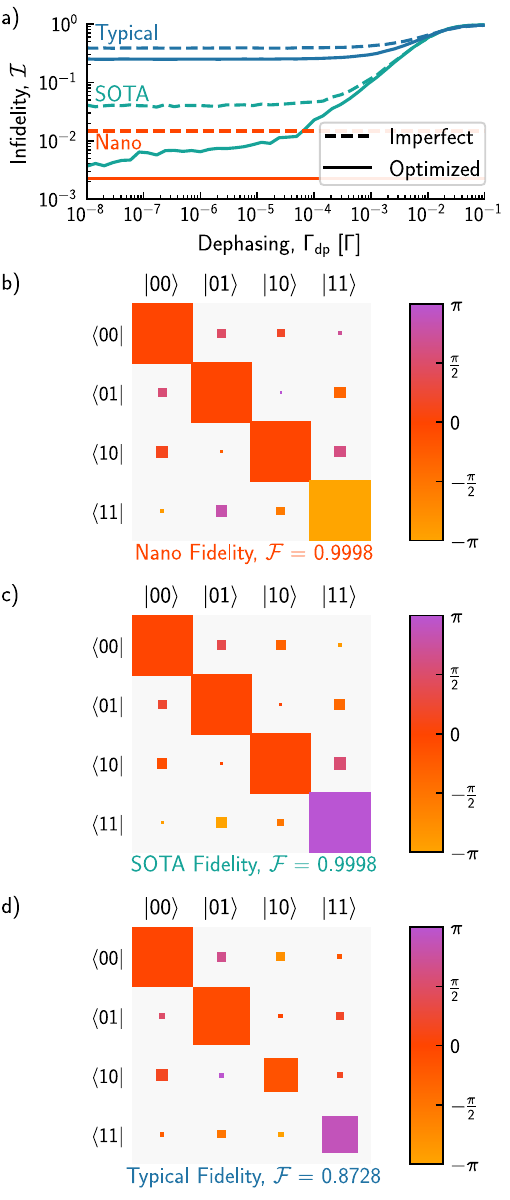}
\caption{\label{fig:CZ} Unheralded CZ gate performance for nanophotonic, state-of-the-art, and typical imperfections (as previously explained in Fig.~\ref{fig:scaling} and Tab.~\ref{Tab:QD_Params}). (a) Imperfect (dashed) and optimized (solid) unitary infidelities ($\mathcal{I}$, c.f. Eq.~\ref{eq:Inf}) as a function of dephasing. All QD parameters other than $\Gamma_\mathrm{dp}$ are as listed in Tab.~\ref{Tab:QD_Params}. Since circuits with only nanophotonic imperfections do not suffer from dephasing, their performance is flat in $\Gamma_\mathrm{dp}$. (b-d) Optimized nanophotonic, state-of-the-art, and typical post-selected $4 \times 4$ computational basis matrices for the unheralded CZ gate, with corresponding conditional output state fidelities listed below each matrix.}
\end{figure}

To demonstrate the possibilities of emitter-based qPICs, we consider the controlled-phase (CZ) gate, which can be used to generate entanglement \cite{CZ_Gate}, and, in the SI, a similar realization of a controlled-not (CNOT) gate that enables universal quantum computation \cite{o2007optical}. A linear-optical unheralded CZ gate can be realized on a $6\times6$ mesh \cite{CZ_Gate}, and in Fig.~\ref{fig:CZ}a we plot the unitary infidelity $\mathcal{I}$ (Eq.~\ref{eq:Inf}) calculated as a function of $\Gamma_{\mathrm{dp}}$ for circuits with only nanophotonic imperfections (red curve, no dephasing), the addition of typical QD imperfections (blue curve), and state-of-the-art QD imperfections (green curve), where for the latter two cases the dephasing varies along the horizontal axis (different from Tab.~\ref{Tab:QD_Params}).
For typical parameters, $\mathcal{I} > 0.25$ regardless of the dephasing rate, consistent with Fig.~\ref{fig:scaling}. Interestingly, the optimized circuit based on typical QD parameters is nearly identical to that based on state-of-the-art when  $\Gamma_{\mathrm{dp}} > 0.01\Gamma$, meaning that in this regime, the effect of dephasing dominates all other emitter parameters. In contrast, the optimized curve for state-of-the-art parameters, including all QD imperfections (green), tends toward that for conventional phase shifters (red) in the limit that dephasing approaches its state-of-the-art value ($\Gamma_\mathrm{dp} \approx 0$). Additionally, while circuits with only nanophotonic imperfections achieve an optimized unitary infidelity of 0.0023, independent of $\Gamma_\mathrm{dp}$, it increases only to 0.0037 at $\Gamma_\mathrm{dp} = 10^{-8}\Gamma$, 0.0065 at $\Gamma_\mathrm{dp} = 10^{-6}\Gamma$, and remains less than 0.01 up to $\Gamma_\mathrm{dp} \sim 10^{-5}\Gamma$. 

More significantly, we consider the fidelity with which logical states are processed by the CZ gate ($\mathcal{F}$) across Figs.~\ref{fig:CZ}b-d. Specifically, here we present the post-selected $4 \times 4$ matrices for the CZ gate in the computational basis for nanophotonic, state-of-the-art, and typical imperfections (as in Tab.~\ref{Tab:QD_Params}). The state fidelity is the chance the CZ gate produces the correct output in the computational basis for any given input, given that the output is in the computational basis. It is shown below the corresponding matrix in each case (state fidelity calculation details are provided in the SI). Although the performance of the typical circuit (Fig.~\ref{fig:CZ}d) significantly differs from the nanophotonic-only (Fig.~\ref{fig:CZ}b), as expected, the circuit based on state-of-the-art QD parameters (Fig.~\ref{fig:CZ}c) achieves $\mathcal{F}=0.9998$, matching the performance of circuits constructed from traditional phase shifters. A similar figure and result is presented in the SI for a CNOT gate. As demonstrated by these results, the state fidelity is typically better than the unitary fidelity, reinforcing the viability of QD-based qPICs.

\section{Conclusions}
Our research demonstrates that qPICs built with reconfigurable QD-phase shifters can perform comparable to those using classical phase shifters. Notably, we show that with state-of-the-art QD parameters, circuits can be scaled up to 10 modes without significant increases in unitary infidelity. This advancement allows QD-based qPICs to efficiently perform operations such as multi-qubit gates, as demonstrated in our study, and to simulate molecular dynamics \cite{ollitrault2021molecular}. The performance of these circuits could be further enhanced by incorporating redundancies, where extra MZIs and phase shifters provide better compensation for imperfections \cite{Pai_Matrix_Opt, Miller:Self_Config, Wilkes:16, Bandyopadhyay:21}. 

However, the practicality of fabricating and managing circuits with numerous emitters remains a topic of inquiry. We recognize that while QDs are the only quantum emitters currently integrated into photonic circuits, alternative technologies based on single organic molecules \cite{turschmann2017chip,toninelli2021single} and defects in diamond \cite{evans2018photon,shandilya2022diamond}, silicon \cite{veldhorst2015two}, or silicon carbide \cite{lukin20204h,lukin2020integrated} are rapidly maturing. Recent experiments with QDs have successfully demonstrated the integration of deterministic QD-based single-photon sources with qPICs \cite{Chanana:22, Kim:17}, and independent control of multiple emitters within a circuit \cite{papon2023independent, Chu:23}, respectively. These advancements, alongside the ability to pre-select specific QDs and integrate them deterministically within a circuit \cite{Pregnolato:20, Thon:09, Schnauber:18, He:17, Liu:17, JinLiu:17}, open the door to larger-scale implementations where emitters would function both as sources and processing elements. Such circuits would be entirely cryogenic, and thus compatible with deterministic sources and detectors, with phase shifters whose operational speeds are determined by emitter lifetimes, potentially enabling GHz rate operation with mild enhancement \cite{pedersen_pin_QD_control, le2022dynamical_Hanna_Nature_Physics}.

%%% Back Matter %%%
\begin{backmatter}

\bmsection{Funding}
Content in the funding section will be generated entirely from details submitted to Prism. Authors may add placeholder text in the manuscript to assess length, but any text added to this section in the manuscript will be replaced during production and will display official funder names along with any grant numbers provided. If additional details about a funder are required, they may be added to the Acknowledgments, even if this duplicates information in the funding section. See the example below in Acknowledgements. For preprint submissions, please include funder names and grant numbers in the manuscript.

\bmsection{Acknowledgments}
The authors thank J. Carolan for generating stimulating discussions on quantum photonic integrated circuits, and gratefully acknowledge support by the Natural Sciences and Engineering Research Council of Canada (NSERC), the Canadian Foundation for Innovation (CFI), the Vector Institute, and Queen's University.

\bmsection{Disclosures}
\noindent The authors declare no conflicts of interest.

\bmsection{Data availability}
Data availability statements are not required for preprint submissions.

\bmsection{Supplemental document}
See Supplement 1 for supporting content. 

\end{backmatter}

%%% References %%%
\bibliography{references}

\end{document}

\maketitle

\section{Transmission and Phase Shift from a Chirally Coupled Quantum Dot}

A quantum dot can be modeled as a two-level system (TLS) with a ground state $|g\rangle$ and an excited state $|e\rangle$, allowing their light-matter interaction to be modeled using the Jaynes-Cummings Hamiltonian \cite{gerry2005introductory_quantum_optics_textbook} as,
\begin{equation}\label{eq:Hamiltonian In waveguide}
\begin{split}
\hat{H}_{S} = -\hbar\Delta_{P}\hat{\sigma}_{eg}\hat{\sigma}_{ge} + \hbar\omega_{P}\hat{f}(r)^{\dagger}\hat{f}(r)-\hat{d}\cdot\hat{E}(r).
\end{split}
\end{equation}
The first term describes the dynamics of the emitter's TLS where $\hat{\sigma}_{ij}=|i\rangle \langle j|$, and $\Delta_{P} = \omega_{P}-\omega_{A}$ is the detuning between the driving light field frequency, $\omega_{P}$, and the emitter's resonant frequency, $\omega_{A}$. The second term, with the bosonic creation and annihilation operators $\hat{f}^{\dagger}$ and $\hat{f}$, describes the energy of the free field at the position $r$. The final term describes the light-matter interaction between the transition dipole of the emitter, $\hat{d}=d^{*}\hat{\sigma}_{eg}+d\hat{\sigma}_{ge}$, and the electric field, $\hat{E} = \hat{E}^{+}+\hat{E}^{-}$. Using the rotating wave approximation and the Lindblad master equation \cite{Elements_Of_Quantum_Optics}, we can determine the steady-state solution for the response of the TLS in terms of its reduced density matrix elements $\rho_{ij}$ as,
\begin{equation}\label{eq:rho_ee and rho_ge}
\begin{split}
\rho_{ee} &= \frac{2\Gamma_{2}\Omega_{P}^{2}}{\Gamma(\Gamma_{2}^{2}+\Delta_{P}^{2}+4(\Gamma_{2}/\Gamma)\Omega_{P}^{2})}\\\rho_{ge}&=\frac{-\Omega_{P}(i\Gamma_{2}+\Delta_{p})}{\Gamma_{2}^{2}+\Delta_{P}^{2}+4(\Gamma_{2}/\Gamma)\Omega_{P}^{2}},
\end{split}
\end{equation}
where $\Gamma = \Gamma_{ee}$ is the rate of spontaneous emission of the TLS, also corresponding to the natural linewidth of the emitter in the frequency spectra. $\Gamma_{2} = \Gamma/2+\Gamma_{dp}$ where $\Gamma_{dp}$ is the rate of dephasing in the system, and $\Omega_{P} = d\cdot E/\hbar$ is the Rabi frequency of the light-matter system.

The electric field operators for the light field can be written in terms of the Green's tensor $\overleftrightarrow{G}(r,r^{\prime})$, the solution for the electric field operator of a point source, and the bosonic creation and annihilation operators \cite{asenjo2017atom_greens_tensor} as,
\begin{equation}\label{eq:Electric field E+ and E-}
\begin{split}
\hat{E}^{+}(r, \omega_{P})&=i\omega_{P}^{2}\mu_{0}\sqrt{\frac{\hbar}{\pi}\epsilon_{0}}\int_{-\infty}^{\infty} dr^{\prime}\sqrt{\epsilon_{I}(r^{\prime},\omega_{P})}\overleftrightarrow{G}(r,r^{\prime})\cdot\hat{f}(r^{\prime},\omega_{P})\\\hat{E}^{-}(r, \omega)&=-i\omega_{P}^{2}\mu_{0}\sqrt{\frac{\hbar}{\pi}\epsilon_{0}}\int_{-\infty}^{\infty} dr^{\prime}\sqrt{\epsilon_{I}(r^{\prime},\omega_{P})}\overleftrightarrow{G}^{*}(r,r^{\prime})\cdot\hat{f}^{\dagger}(r^{\prime},\omega_{P}).
\end{split}
\end{equation}

Equations \ref{eq:Hamiltonian In waveguide} and \ref{eq:Electric field E+ and E-} allow us to write the time evolution of the bosonic annihilation operator using the Heisenberg Equation of motion as,
\begin{equation}\label{eq:Heisenberg Equation of Motion}
\begin{split}
\dot{\hat{f}}(r,\omega_{P})&=\frac{i}{\hbar}\left[\hat{H},\hat{f}(r,\omega_{P})\right]\\
&=-i\omega_{P}\hat{f}(r,\omega_{P})+\omega_{P}^{2}\mu_{0}\sqrt{\frac{\hbar}{\pi}\epsilon_{0}}d(r_{A})\sqrt{\epsilon_{I}(r,\omega_{P})}\overleftrightarrow{G}^{*}(r_{A},r,\omega_{P})\hat{\sigma}_{ge},
\end{split}
\end{equation}
where $r_{A}$ is the position of the emitter in the waveguide. Formally integrating this equation from time $t^{\prime}$ to $t$ results in \cite{dung2002resonant_formal_integral_ref},
\begin{equation}\label{eq:Integrated solution to bosonic operator from equation of motion}
\begin{split}
\hat{f}(r,\omega_{P},t)&=\hat{f}(r,\omega_{P},t^{\prime})\exp(-i\omega_{P}(t-t^{\prime}))+\\&\omega_{P}^{2}\mu_{0}\sqrt{\frac{\hbar}{\pi}\epsilon_{0}}\int_{0}^{t} dt^{\prime}d(r_{A})\sqrt{\epsilon_{I}(r,\omega_{P})}\overleftrightarrow{G}^{*}(r_{A},r,\omega_{P})\hat{\sigma}_{ge}(t^{\prime})e^{-i\omega_{P}(t-t^{\prime})},
\end{split}
\end{equation}
where the first term describes a free excitation in the system that does not interact with the emitter, and the second term describes an interaction with the emitter either through a decay from excited to ground state or through a scattering event (virtual transition). Thus, we can write the electric field operator as,
\begin{equation}\label{eq:E field breakdown into incident and scattered}
\begin{split}
\hat{E}^{+} = \hat{E}_{P}^{+} + \hat{E}_{S}^{+},
\end{split}
\end{equation}
where $\hat{E}_{P}^{+}$ represents the incident field and $\hat{E}_{S}^{+}$ represents the scattered field. The transmission coefficient for the light in the system is written in terms of the expectation values of these fields as,
\begin{equation}\label{eq:transmission coefficient}
\begin{split}
t = \frac{\langle\hat{E}^{+}\rangle}{\langle\hat{E}^{+}_{P}\rangle},
\end{split}
\end{equation}
and the phase shift on the light from the interaction with the emitter is,
\begin{equation}\label{eq:phase shift from t}
\begin{split}
\phi = \arg(t).
\end{split}
\end{equation}

Equations \ref{eq:Electric field E+ and E-} and \ref{eq:Integrated solution to bosonic operator from equation of motion}, with the help of the Green's tensor identity \cite{asenjo2017atom_greens_tensor} and Kramer's Kronig relations \cite{dzsotjan2011dipole_Kramers_Kronig}, allow the scattered field in terms of the incident field as,
\begin{equation}\label{eq:Scattered field after Kramers-Kronig}
\begin{split}
\hat{E}_{S}^{+}(r,t)&=\frac{1}{\hat{\Omega}_{P}}g(r,r_{A}, \omega_{A})\hat{\sigma}_{ge}(t)\hat{E}_{P}^{+}(r_{A},t),
\end{split}
\end{equation}
where we have defined the Rabi frequency operator as $\hat{\Omega}_{P} = d^{*}\cdot \hat{E}_{P}^{+}/\hbar$, where $\langle\hat{\Omega}_{P}\rangle =\Omega_{P}$, and we use the dipole-projected Green's function \cite{asenjo2017atom_greens_tensor},
\begin{equation}\label{eq:dipole-projected Green's function}
\begin{split}
g(r_{i},r_{j},\omega) &= \frac{\mu_{0}\omega^{2}}{\hbar}d^{*}(r_{i})\cdot G(r_{i},r_{j},\omega)\cdot d(r_{j}).
\end{split}
\end{equation}

For a chirally coupled quantum dot, the dipole projected Green's function becomes,
\begin{equation}\label{eq:chiral dipole-projected Green's function}
\begin{split}
g(r,r_{A},\omega) &= i\Gamma(\Theta(r_{A} - r)\beta_{L} + \Theta(r - r_{A})\beta_{R})e^{ik_{p}|r-r_{A}|},
\end{split}
\end{equation}
where $\Theta$ is the Heaviside function and we define the couplings for photons moving left $\beta_{L}$ and for photons moving right $\beta_{R}$ as,
\begin{equation}\label{eq:beta L}
\begin{split}
\beta_{L} = \frac{\Gamma_{L}}{\Gamma}=\frac{\Gamma_{L}}{\Gamma_{L} + \Gamma_{R} + \Gamma_{Loss}},
\end{split}
\end{equation}
\begin{equation}\label{eq:beta R}
\begin{split}
\beta_{R} = \frac{\Gamma_{R}}{\Gamma}=\frac{\Gamma_{R}}{\Gamma_{L} + \Gamma_{R} + \Gamma_{Loss}}.
\end{split}
\end{equation}
In the low-power regime (weak coherent beam) where $\Omega_{P} \to 0$, the scattering event that imparts a phase shift will dominate over spontaneous emission, allowing the emitter to act as a phase shifter. From Equations \ref{eq:rho_ee and rho_ge} and \ref{eq:transmission coefficient} the transmission coefficient becomes,
\begin{equation}\label{eq:t}
\begin{split}
t&=1-\Gamma\beta_{\mathrm{R}}\frac{\Gamma_{2}+i\Delta_{\mathrm{p}}}{\Gamma_{2}^{2}+\Delta_{\mathrm{p}}^{2}},
\end{split}
\end{equation}
where we have defined the forward direction as right for simplicity. The phase shift can then be calculated with Equation \ref{eq:phase shift from t}. With the same assumptions, the observable transmission, $T$, can be written as,
\begin{equation}\label{eq:transmission observable}
\begin{split}
T &= \frac{\langle \hat{E}^{-}\hat{E}^{+}\rangle}{\langle \hat{E}^{-}_{P}\hat{E}^{+}_{P}\rangle}\\
T&=1-2\Gamma\Gamma_{2}\frac{\beta_{\mathrm{R}}(1-\beta_{\mathrm{R}})}{(\Gamma_{2}^{2}+\Delta_{\mathrm{p}}^{2})}.
\end{split}
\end{equation}

\section{Ideal Phase Solution Using Unitary Decomposition}\label{sec:MZI Meshes and Decomposition}

To solve the ideal phase shifts for a target unitary we follow the decomposition and recombination method proposed by Clements et al. \cite{clements2016optimal} where an ideal MZI is built up from the $2 \times 2$ transfer matrices of two $50:50$ directional couplers and two phase shifters ($\phi$, $2\theta$) as,
\begin{equation}\label{eq:Our MZI}
\begin{split}
MZI &=\frac{1}{\sqrt{2}}\begin{bmatrix}1 & i\\
i & 1
\end{bmatrix}\begin{bmatrix}e^{i2\theta} & 0\\
0 & 1
\end{bmatrix}\frac{1}{\sqrt{2}}\begin{bmatrix}1 & i\\
i & 1
\end{bmatrix}\begin{bmatrix}e^{i\phi} & 0\\
0 & 1
\end{bmatrix}\\&=ie^{i\theta}\begin{bmatrix}e^{i\phi}sin\theta & cos\theta\\
e^{i\phi}cos\theta & -sin\theta
\end{bmatrix},
\end{split}
\end{equation}
which can be expanded into the $N \times N$ matrix,
\begin{equation}\label{eq:T_mn}
T_{m,n}^{(p)} = \begin{bmatrix}1 & 0 & \dots & \dots & \dots & 0\\
0 & \ddots & \dots & \dots & \dots & \vdots\\
\vdots & \dots & ie^{i\theta}e^{i\phi}sin\theta & ie^{i\theta}cos\theta & \dots & \vdots\\
\vdots & \dots & ie^{i\theta}e^{i\phi}cos\theta & -ie^{i\theta}sin\theta & \dots & \vdots\\
\vdots & \dots & \dots & \dots & \ddots & 0\\
0 & \dots & \dots & \dots & 0 & 1\\
\end{bmatrix},
\end{equation}
representing the $pth$ MZI between modes $m$ and $n$ in the circuit. To solve for the ideal phases the Clements method \cite{clements2016optimal} then decomposes the unitary into a diagonal matrix $D$ by applying MZIs/inverse MZIs using each of the $N(N-1)/2$ MZIs to nullify an off-diagonal entry. The result for a $4 \times 4$ unitary is,
\begin{equation}\label{eq:Clements Decomposition}
\begin{split}
\Tilde{T}_{3,4}^{(1)}\Tilde{T}_{2,3}^{(1)}UT_{1,2}^{(0)-1}T_{3,4}^{(0)-1}T_{2,3}^{(0)-1}T_{1,2}^{(1)-1} &= D,
\end{split}
\end{equation}
where the $~$ indicates that these MZIs do not correspond to the hardware MZIs as they will be changed during the recombination steps. To determine the required phases for these MZIs, the algorithm is as follows; If a $T_{m,n}^{(p)-1}$ is being applied, the phases are chosen as,
\begin{equation}\label{eq:LHS Phases}
\begin{split}
\theta &= \frac{\pi}{2} - \arctan\left(\left|\frac{U_{Null}[x,m]}{U_{Null}[x,n]}\right|\right)\\
\phi &= \pi + \arg\left(\frac{U_{Null}[x,m]}{U_{Null}[x,n]}\right),
\end{split}
\end{equation}
where $U_{Null}$ corresponds to the partially decomposed matrix at the current decomposition step, and the index $x$ is the row index of the element being nullified. Conversely, if a $T_{m,n}^{(p)}$ is being applied to the LHS, the phases are chosen as,
\begin{equation}\label{eq:RHS Phases}
\begin{split}
\theta &= \frac{\pi}{2} - \arctan\left(\left|\frac{U_{Null}[n,y]}{U_{Null}[m,y]}\right|\right)\\
\phi &= \pi + \arg\left(\frac{U_{Null}[n,y]}{U_{Null}[m,y]}\right),
\end{split}
\end{equation}
where now the column index $y$ corresponds to the column of the element being nullified.

For the recombination step, isolate for $U$ in Equation \ref{eq:Clements Decomposition}, and shift the diagonal matrix to the left, replacing it and the $\tilde{T}_{m,n}^{(p)-1}$'s according to,
\begin{equation}\label{eq:Diagonal Matrix Equivalency}
\begin{split}
\Tilde{T}_{m,n}^{(p)-1}D = D_{1}T_{m,n}^{(p)},
\end{split}
\end{equation}
where $D_{1}$ is another diagonal matrix and $T_{m,n}^{(p)}$ is the MZI that corresponds to the hardware implementation. The new phases for $T_{m,n}^{(p)}$, following the Clements solution are,
\begin{equation}\label{eq:New Phases for Shifting D}
\begin{split}
\theta &= \frac{\pi}{2} - \arctan\left(\left|\frac{M[n,m]}{M[n,n]}\right|\right)\\
\phi &= \pi + \angle\frac{M[n,m]}{M[n,n]},
\end{split}
\end{equation}
where $M = \Tilde{T}_{m,n,p}^{-1}D$. From here, we determine the new diagonal matrix as,
\begin{equation}\label{eq:New Diagonal Matrix}
\begin{split}
D_{1} = \Tilde{T}_{m,n}^{(p)-1}DT_{m,n}^{(p)-1}.
\end{split}
\end{equation}
By repeating these steps until D is on the left side of the formula, we get, for a $4 \times 4$, an equation for the ideal unitary as,
\begin{equation}\label{eq:Unitary Equation for 4x4}
\begin{split}
U &= D^{\prime}T_{2,3}^{(1)}T_{3,4}^{(1)}T_{1,2}^{(1)}T_{2,3}^{(0)}T_{3,4}^{(0)}T_{1,2}^{(0)},
\end{split}
\end{equation}
where we have relabelled the final diagonal matrix as $D^{\prime}$, which corresponds to global phase shifts applied at the end of the MZI mesh.

\section{Imperfect Transfer Matrix Generation}
Imperfections are added to the transfer matrices at the MZI level, which combine to form an imperfect transfer matrix using Equation \ref{eq:Unitary Equation for 4x4}. The first imperfection we consider is nanophotonic loss, originating from waveguide propagation losses and beam splitter losses. For this, we use the common \cite{bogaerts_programmable_2020, jacob_ewaniuk_imperfect_2023, Shokraneh_mesh, shokraneh_theoretical_2021} assumption that loss is balanced across MZI paths, modeling the loss as,
\begin{equation}\label{eq:Nanophotonic Loss}
Loss = \begin{bmatrix}\sqrt{1-L} & 0\\
0 & \sqrt{1-L}
\end{bmatrix}.
\end{equation}
Beam splitter error on the directional couplers is also included, modeled using the standard variable beam splitter matrix,
\begin{equation}\label{eq:Imperfect BS}
BS = \begin{bmatrix}\sqrt{r} & i\sqrt{1-r}\\
i\sqrt{1-r} & \sqrt{r}
\end{bmatrix},
\end{equation}
where $r$ is the reflectivity of the beam splitter and $r = 0.5$ in the ideal case.

Quantum dot phase shift imperfections are factored into the model as well. Imperfect coupling, $\beta = \beta_{L} + \beta_{R} < 1$, and imperfect directionality, $D = (\beta_{R} - \beta_{L})/(\beta_{L} + \beta_{R}) < 1$, are included through their effects on quantum dot loss, modeled as $\gamma = 1 - T$, where $T$ is the observable transmission in the system calculated using Equation \ref{eq:transmission observable}. This is included using the lossy phase shift matrix,
\begin{equation}\label{eq:Imperfect PS}
\begin{split}
PS = \begin{bmatrix}\sqrt{1-\gamma}e^{\phi} & 0\\
0 & 1
\end{bmatrix}.
\end{split}
\end{equation}

\begin{figure}[ht]
\centering
\includegraphics[scale=0.31]{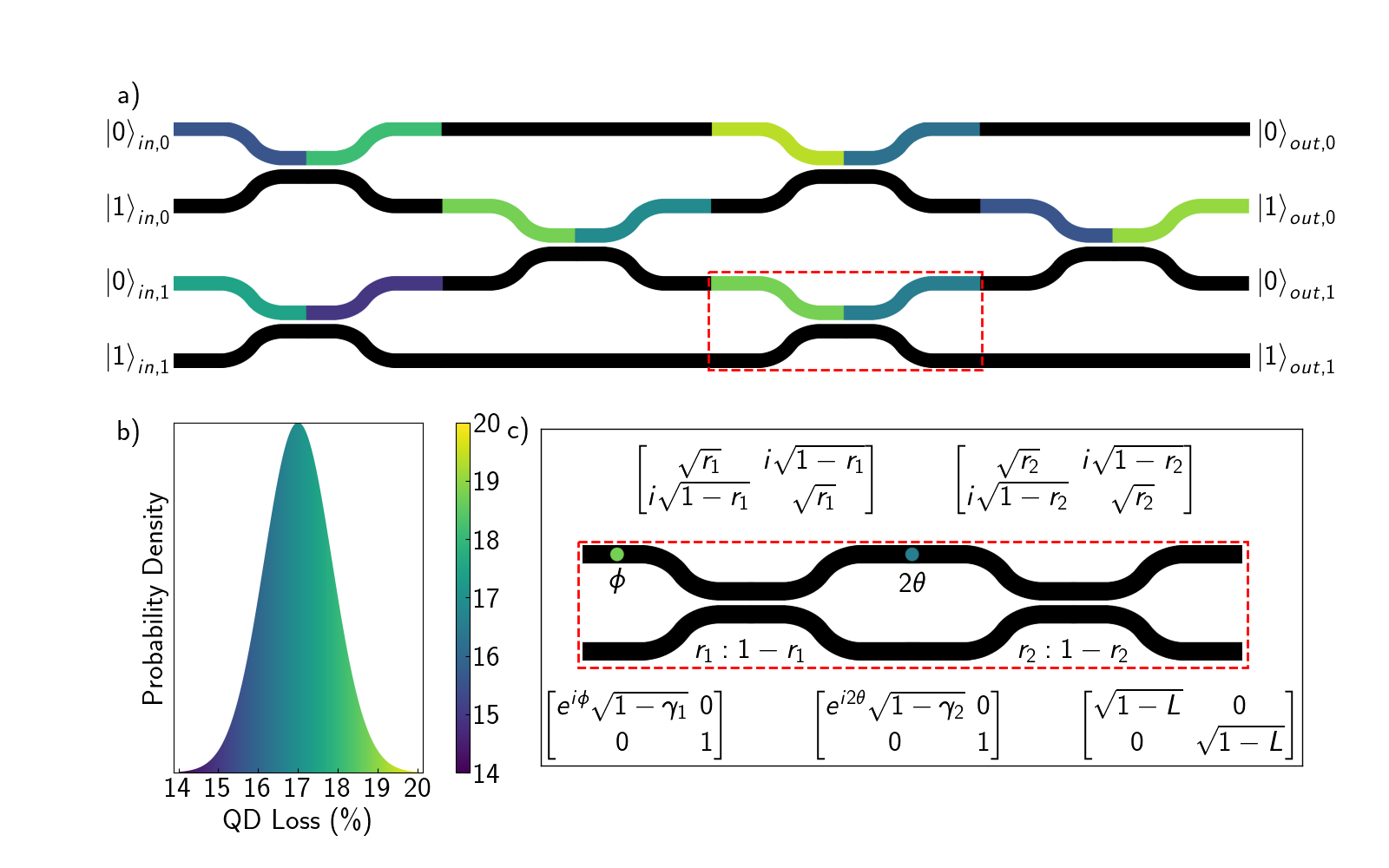}
\caption{Imperfect circuit transfer matrix generation with example quantum dot loss distribution. (a) $4 \times 4$ MZI mesh circuit depicting QD loss for each quantum dot in the MZIs. The color in the top-left branch of each MZI indicates the loss for the $\phi$ phase shifter, and the color in the top-right branch of each MZI indicates the loss for the $2\theta$ phase shifter. If the path is black this indicates no QD loss in that region of the circuit. (b) Normal distribution for QD loss with a central value of $17\%$ and a standard deviation of $5\%$ of $17\%$. (c) MZI with transfer matrices for the two QD phase shifters with losses $\gamma_{1}$ and $\gamma_{2}$ as colored, the two beam splitters with reflectivities $r_{1}$ and $r_{2}$ and nanophotonic loss per MZI of $L$.} 
\label{fig:QDLossDist}
\end{figure}

Figure \ref{fig:QDLossDist} depicts how the quantum dot loss $\gamma$ is sampled for each phase shifter from a distribution with a central value $\gamma = 1 - T$ with a standard deviation of $5\%$ of the central value to model fluctuations between different quantum dots. Combining all these imperfections into the model, each MZI is calculated using the imperfect matrix,
\begin{equation}\label{eq:Imperfect MZI}
\begin{split}
MZI&=\begin{bmatrix}\sqrt{r_{2}} & i\sqrt{1-r_{2}}\\
i\sqrt{1-r_{2}} & \sqrt{r_{2}}
\end{bmatrix}\begin{bmatrix}\sqrt{1-\gamma_{2}}e^{i2\theta} & 0\\
0 & 1
\end{bmatrix}\begin{bmatrix}\sqrt{r_{1}} & i\sqrt{1-r_{1}}\\
i\sqrt{1-r_{1}} & \sqrt{r_{1}}
\end{bmatrix}\\&\begin{bmatrix}\sqrt{1-\gamma_{1}}e^{i\phi} & 0\\
0 & 1
\end{bmatrix}\begin{bmatrix}\sqrt{1-L} & 0\\
0 & \sqrt{1-L}
\end{bmatrix},
\end{split}
\end{equation}
which is then combined with all other imperfect MZIs to build an $N \times N$ imperfect transfer matrix. This is depicted as a circuit in Figure \ref{fig:QDLossDist}a, which shows how MZIs combine to form a $4 \times 4$ circuit, where Figure \ref{fig:QDLossDist}c shows the breakdown of each MZI component with its appropriate transfer matrix.

\begin{figure}[ht]
\centering
\includegraphics[scale=0.4]{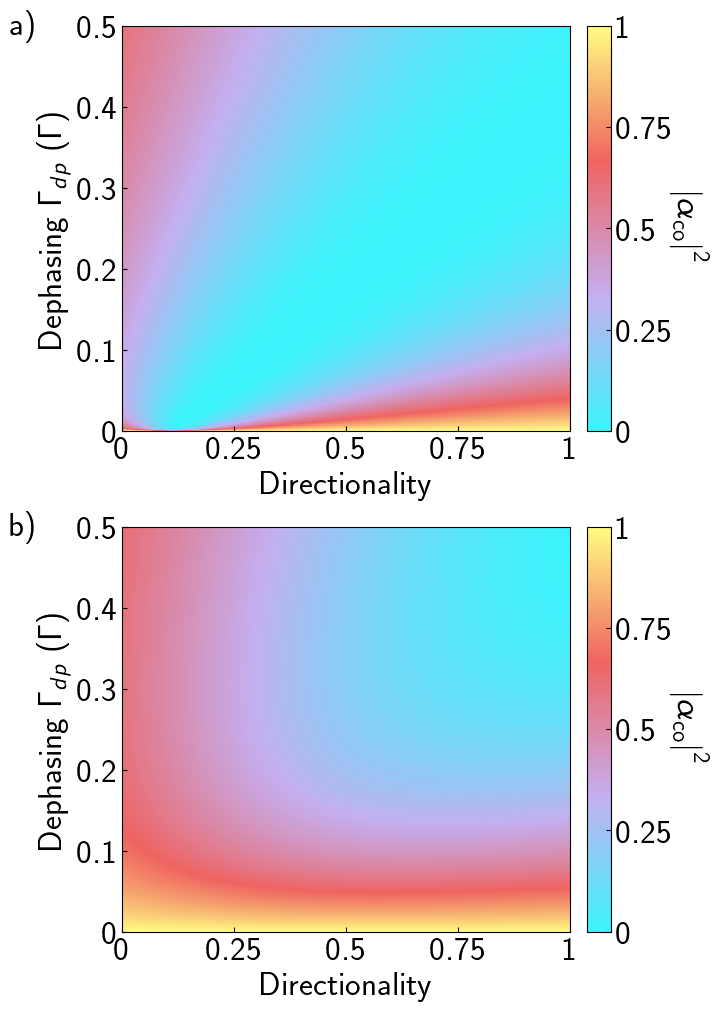}
\caption{Coherent phase shift probability ($\left|\alpha_{\mathrm{co}}\right|^2$) maps spanning dephasing ($0\Gamma$ to $0.5\Gamma$) and directionality ($0$ to $1$) with coupling of  $\beta = 0.9$ (a) $\Delta_{\mathrm{P}} = 0\Gamma$ (b) $\Delta_{\mathrm{P}} = 0.3\Gamma$.} 
\label{fig:DephCoherentProb}
\end{figure}

This imperfect transfer matrix does not include dephasing and spectral diffusion, which are included in QD imperfections. Dephasing provides a chance of having an incoherent interaction with an average phase shift of $0$, calculated as $\left|\alpha_{\mathrm{inc}}\right|^2 = (T - |t|^{2})/T$. Consequently, the probability of a coherent interaction is $\left|\alpha_{\mathrm{co}}\right|^2 = 1 - \left|\alpha_{\mathrm{inc}}\right|^2$. Figure \ref{fig:DephCoherentProb} depicts the coherence probability based on dephasing and directionality for detunings of  $\Delta_{P} = 0\Gamma$ in Figure \ref{fig:DephCoherentProb}a and $\Delta_{P} = 0.3\Gamma$ in Figure \ref{fig:DephCoherentProb}. The larger the detuning, the higher the coherence probability, since there is less interaction with the quantum dot and thus less dependence on dephasing.

\begin{figure}[ht]
\centering
\includegraphics[scale=0.45]{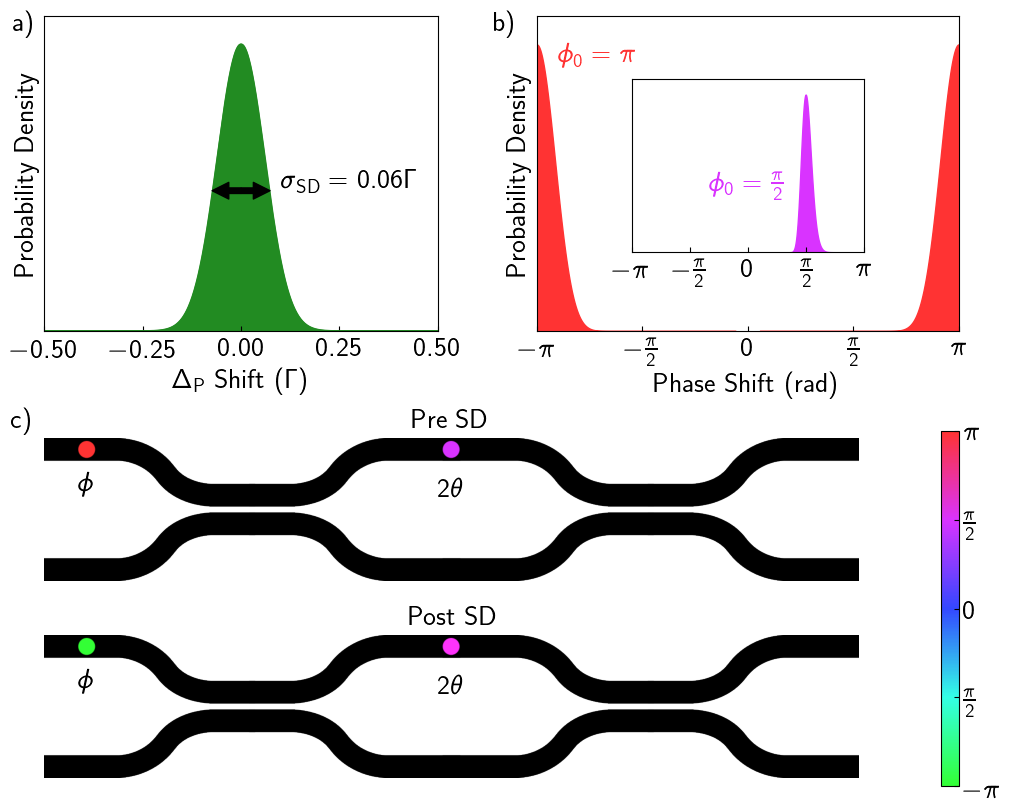}
\caption{Spectral Diffusion effects on detuning and phase shift for $\sigma_{SD} = 0.06\Gamma$. (a) Normal distribution for detuning shift from spectral diffusion. (b) Phase shift distribution example for initial phases of $\pi$ and $\pi/2$ (inset). (c) Example MZI with initial phases of $\phi = \pi$ and $2\theta = \pi/2$ before and after sampling spectral diffusion.} 
\label{fig:SD_Sup}
\end{figure}

Dephasing is modeled using a Monte-Carlo simulation with $500$ samples, where in each trial each phase shifter is sampled to be on/off based on its QD's incoherence probability, generating $500$ imperfect unitaries on which we average the performance.

Spectral diffusion is also included in these Monte-Carlo simulations, where for each phase shifter we sample a shift in detuning, and consequently phase, from a normal distribution with a standard deviation of $\sigma_{\mathrm{SD}}$, modeling the fluctuations from spectral diffusion in the system. Figure \ref{fig:SD_Sup} depicts the spectral diffusion effects on detuning and phase shift for an example spectral diffusion of $\sigma_{\mathrm{SD}} = 0.06\Gamma$. The detuning shift is sampled from the normal distribution shown in Figure \ref{fig:SD_Sup}a, with Figure \ref{fig:SD_Sup}b showing the phase shift distribution for phase shifts of $\pi$ and $\pi/2$, and Figure \ref{fig:SD_Sup}c showing a sample MZI pre and post spectral diffusion shift. It is evident that large phase shifts (like $\pi$) where $\Delta_{P} \to 0$ will have large fluctuations since detuning will often change signs, whereas smaller phase shifts with larger detuning such as $\pi/2$ will not see as much spectral diffusion influence.

\section{Circuit Accuracy Measurement Methods and Phase Optimization}\label{sec:circuitAcc}
The accuracy of an imperfect transfer matrix (non-ideal U) for each circuit is calculated using the matrix infidelity \cite{clements2016optimal},
\begin{equation}\label{eq:Inf}
\begin{split}
\mathcal{I} &= 1 - \left|\frac{\mathrm{tr}(\mathrm{U}^{\dagger}\mathrm{U}_{\mathrm{non}})}{\sqrt{N \mathrm{tr}(\mathrm{U}_{\mathrm{non}}^{\dagger}\mathrm{U}_{\mathrm{non}})}}\right|^{2} 
\end{split},
\end{equation}
which excludes balanced losses, allowing us to focus on accuracy over count rate. However, for the CZ and CNOT gate, we consider accuracy using the post-selected output state infidelity instead. Since the CNOT and CZ are two-photon gates, their matrices must be converted into the fock basis by calculating the matrix permanent \cite{AA_Protocol}. However, only four inputs/outputs of these expanded matrices correspond to the computational two-qubit basis states $\{|00\rangle,|01\rangle,|10\rangle,|11\rangle\}$. Thus, for each input-output pair, we calculate the unconditional output state following the equation,
\begin{equation}\label{eq:Output Unconditional Fidelity}
\begin{split}
|\psi_{\mathrm{out,non}}^{(i,\mathrm{unc})}\rangle &= U_{\mathrm{non}}|\psi_{\mathrm{in}}^{(i)}\rangle,
\end{split}
\end{equation}
where $|\psi_{\mathrm{out,non}}^{(i,\mathrm{unc})}\rangle$ is the imperfect unconditional output state in the fock basis, and $U_{\mathrm{non}}$, $\psi_{\mathrm{in}}^{(i)}$ are also in the fock basis. These states are then post-selected by truncating them down to the computational basis states resulting in four-entry vectors and re-normalizing them. The conditional output state infidelity for an input-output pair is then calculated using,
\begin{equation}\label{eq:Output Conditional infidelity}
\begin{split}
\mathcal{I}_{i}^{(\mathrm{con})} &= 1 - \left|\left\langle \psi_{\mathrm{out}}^{(i,\mathrm{con})}|\psi_{\mathrm{out,non}}^{(i,\mathrm{con})} \right\rangle\right|^{2}.
\end{split}
\end{equation}

We also consider the post-selected $4 \times 4$ computational basis matrices in the main text, which are found by truncating the fock basis matrix down to the $4 \times 4$ matrix corresponding to the four computational basis inputs and outputs and re-normalizing the matrix.

To perform phase shift optimization on these circuits, we use the appropriate cost function (Equation \ref{eq:Inf} or \ref{eq:Output Conditional infidelity}) to measure the circuit error, and perform an optimization on all phases using the BOBYQA optimization algorithm \cite{powell2009bobyqa}. To do this, we began by determining the phase shifts for a given unitary in the ideal case using the Clements decomposition method outlined in Section \ref{sec:MZI Meshes and Decomposition}. Then, imperfections were added to the circuit to calculate the imperfect transfer matrix and its associated infidelity. For optimization, the ideal phases were chosen as initial phase guesses, with phase constraints of $[-\pi,\pi]$, which are required as this is a constrained optimization algorithm. When optimizing circuits with dephasing/spectral diffusion, the cost function averages the infidelity for 500 sampled matrices to find the cost for every optimization step.

\section{CNOT Gate Results}
\begin{figure}[htbp]
\centering
\includegraphics[scale=0.35]{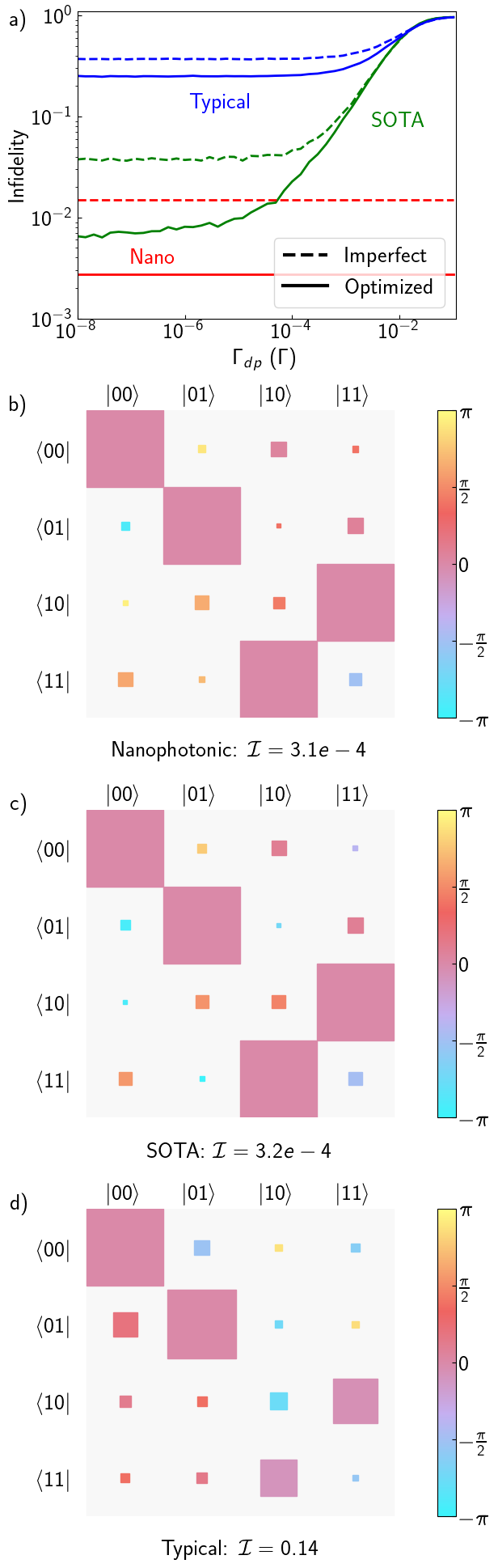}
\caption{Unheralded CNOT performance for nanophotonic, state-of-the-art, and typical imperfections. (a) Matrix infidelity as a function of dephasing. The nanophotonic infidelities with no dephasing are shown as horizontal lines across the figure.  (b-d) Optimized nanophotonic, state-of-the-art, and typical post-selected $4 \times 4$ computational basis matrices for the unheralded CNOT gate, with conditional output state fidelities listed.} 
\label{fig:CNOT}
\end{figure}

Similar to the results for the CZ gate in the main text, here we consider the unheralded CNOT gate, a 6-mode circuit with $1/9$ probability of success. The performance was first measured using the matrix infidelity (Equation \ref{eq:Inf}) spanning from dephasing of $\Gamma_{dp} = 10^{-8}\Gamma \to 10^{-1}\Gamma$ for nanophotonic, state-of-the-art and typical imperfections as shown in Figure \ref{fig:CNOT}. These imperfections are the same as listed in the main text. The results show that performance and optimization depend heavily on dephasing, where incoherent interactions heavily hinder performance. However, state-of-the-art quantum dot imperfections can be optimized to perform on par with nanophotonic imperfections with no dephasing. Next, as described in Section \ref{sec:circuitAcc}, the CNOT gate circuit was optimized based on the average conditional output state infidelity (Equation \ref{eq:Output Conditional infidelity}). The optimized post-selected $4 \times 4$ computational basis matrices for nanophotonic, state-of-the-art, and typical imperfections are shown in Figure \ref{fig:CNOT}b-d, along with their optimized conditional output state infidelities. The output state performance is near perfect for nanophotonic and state-of-the-art imperfections, despite overall performing poorer than the CZ due to increased circuit complexity.

\section{Random Circuit Infidelity Distributions}\label{sec:Beta Distributions}
\begin{figure}[htbp]
\centering
\includegraphics[scale=0.6]{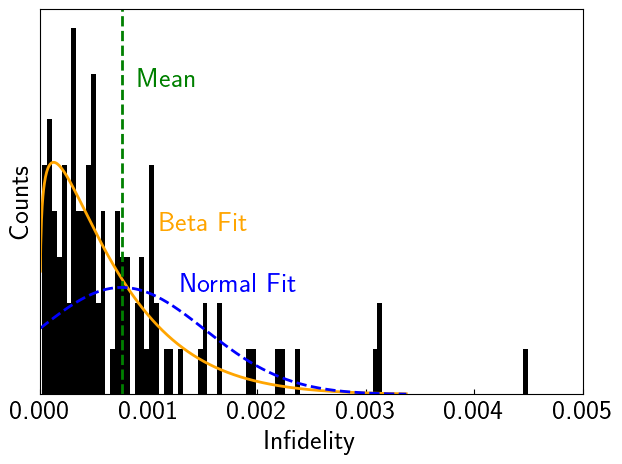}
\caption{Matrix infidelity histogram over 100 Haar random $N = 4$ circuits for a beam splitter error of $2.3\%$, with no other imperfections. Here we show the mean of the data (green), a beta fit (orange), and a normal fit (blue), to the data.} 
\label{fig:Beta Distribution}
\end{figure}

Each infidelity data point in the results is the result of a beta distribution across 100 Haar random unitary matrices to simulate random circuit performance accurately. A beta distribution is used to accurately average the data as the infidelity distribution across 100 samples is asymmetric and includes outliers. A beta distribution follows the probability density function,
\begin{equation}\label{eq:Beta Prob Density}
\begin{split}
f(x) = \frac{(x-a)^{p-1}(b-x)^{q-1}}{B(p,q)(b-a)^{p+q-1}},
\end{split}
\end{equation}
where $a$ and $b$ are the lower and upper bounds on $x$, $p$ and $q$ are shape parameters where $p,q > 0$ and $B(p,q)$ is the beta function that follows the equation,
\begin{equation}\label{eq:Beta Function}
\begin{split}
B(\alpha,\beta) = \int^{1}_{0} t^{\alpha - 1}(1-t)^{\beta-1}dt.
\end{split}
\end{equation}
Figure \ref{fig:Beta Distribution} shows an example infidelity histogram for an $N = 4$ circuit with an arbitrary choice of $2.3\%$ beam splitter error. This figure also plots a beta fit, normal distribution fit, and the mean of the data.
The mean and normal fit in Figure \ref{fig:Beta Distribution} both result in average infidelity that is higher than the majority of the data. Conversely, the beta fit accounts for the asymmetry, representing the data accurately. Thus, the mean of the beta fit is used for random circuit infidelity results.

\section{Linear Response Pulse Width Constraint}
The scattering of few-photon Fock states by a two-level system chirally-coupled to a one-dimensional waveguide can be described using an input-output formalism, as demonstrated in \cite{fan2010input}. Here, we will use the scattering matrices for one and two-photon Fock states to demonstrate the linear phase response that arises when the photon pulse width $\sigma_\mathrm{p}$ is much less than the linewidth of the quantum emitter $\Gamma$. Then, we select a practical constraint for $\sigma_\mathrm{p}$ where the emitter can be used as a linear phase shifter for Fock states of up to two photons. For simplicity, we will assume perfect directionality and no loss.

As shown in \cite{fan2010input}, the single-photon scattering matrix elements are given by,
\begin{equation} \label{eq:scatteringmatrix1}
    \left\langle p \right| S \left| q\right\rangle = t(q)\delta(q - p) \quad:\quad t(q) \equiv \frac{q - \omega_A - \frac{i\Gamma}{2}}{q - \omega_A + \frac{i\Gamma}{2}}, 
\end{equation}
where $S$ is the scattering matrix, $\left|q\right\rangle = a^\dag_q\left|0\right\rangle$, and $\omega_A$ is the transition frequency of the quantum emitter. Consider a single-photon wave packet with a pulse shape $\alpha(\omega)$, centered at angular frequency $\omega_0$, as the input state,
\begin{equation} \label{eq:input1}
    \left|\mathrm{in}\right\rangle_1 = \int_{-\infty}^\infty d\omega \alpha(\omega)a_\omega^\dag\left|0\right\rangle.
\end{equation}
The output state can be derived as,
\begin{equation}
    \left|\mathrm{out}\right\rangle_1 = \int_{-\infty}^\infty d\omega t(\omega)\alpha(\omega)a_\omega^\dag\left|0\right\rangle,
\end{equation}
by first applying the scattering matrix to the input, then inserting closure followed by Eq.~\ref{eq:scatteringmatrix1}. Most commonly, $\alpha(\omega)$ will take either a Gaussian or Lorentzian form, both of which can be parameterized by width $\sigma_\mathrm{p}$. For either form, $\alpha(\omega) \sim \delta(\omega - \omega_0)$ in the limit $\sigma_\mathrm{p} \to 0$ which implies,
\begin{equation}
    \left|\mathrm{out}\right\rangle_1 \sim t(\omega_0)a_{\omega_0}^\dag\left|0\right\rangle.
\end{equation}
In the lossless case, $\left|t(\omega)\right| = 1$ and $\mathrm{arg}\left\{t(\omega)\right\}$ varies with detuning $\Delta = \omega_A - \omega_0$ from $-\pi$ to $\pi$. Therefore, the quantum emitter acts as a perfect phase shifter when acting on single-photon Fock states in the monochromatic limit.

\begin{figure}[htbp]
\centering
\includegraphics[width=.6\linewidth]{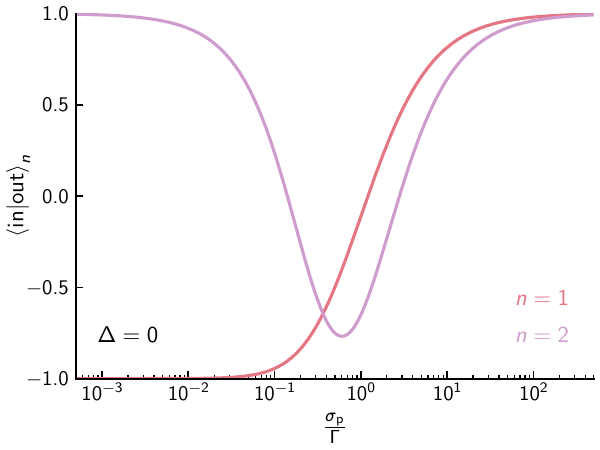}
\caption{Overlap between an input uncorrelated $n$-photon wave packet with Lorentzian pulse shape and the output state achieved after the photons scatter from a perfectly chiral two-level quantum emitter coupled to a 1D waveguide, as a function of pulse width $\sigma_\mathrm{p}$. Each photon is assumed to be centered on resonance with the quantum emitter transition frequency such that $\Delta = \omega_A - \omega_0 = 0$, and the interaction is assumed to be lossless.}
\label{fig:fockscattering}
\end{figure}

We now turn to the two-photon scattering matrix elements, derived in Ref.~\cite{fan2010input} as,
\begin{align} \label{eq:scatteringmatrix2}
    \left\langle p_1p_2\right|S\left|q_1q_2\right\rangle =\quad &t(p_1)t(p_2)\left[\delta(q_1 - p_1)\delta(q_2 - p_2) + \delta(q_1 - p_2)\delta(q_2 - p_1)\right]\nonumber \\ &+ \frac{i\sqrt{\Gamma}}{\pi}s(p_1)s(p_2)\left[s(q_1) + s(q_2)\right]\delta(q_1 + q_2 - p_1 - p_2),
\end{align}
where $\left|q_1q_2\right\rangle = \frac{1}{\sqrt{2}}a^\dag_{q1}a^\dag_{q2}\left|0\right\rangle$ and $s(\omega)$ is a measure of the excitation of the emitter by a single-photon wave packet, defined by,
\begin{equation}
    s(\omega) \equiv \frac{\sqrt{\Gamma}}{q - \omega_A + \frac{i\Gamma}{2}}.
\end{equation}
With the input as the two-photon analog of Eq.~\ref{eq:input1},
\begin{equation}
    \left|\mathrm{in}\right\rangle_2 = \frac{1}{\sqrt{2}}\int_{-\infty}^\infty d\omega_1\int_{-\infty}^\infty d\omega_2 \alpha(\omega_1)\alpha(\omega_2)a_{\omega_1}^\dag a_{\omega_2}^\dag\left|0\right\rangle,
\end{equation}
the output can be derived by following the same procedure to obtain the result,
\begin{align}
    \left|\mathrm{out}\right\rangle_2 = &\frac{1}{\sqrt{2}}\int_{-\infty}^\infty d\omega_1\int_{-\infty}^\infty d\omega_2 \Bigg[t(\omega_1)t(\omega_2)\alpha(\omega_1)\alpha(\omega_2) + \frac{i\sqrt{\Gamma}}{2\pi}s(\omega_1)s(\omega_2)\nonumber \\ &\times \int_{-\infty}^\infty dp \alpha(\omega_1 + \omega_2 - p)\alpha(p)\left(s(\omega_1 + \omega_2 - p) + s(p)\right)\Bigg]a_{\omega_1}^\dag a_{\omega_2}^\dag\left|0\right\rangle.
\end{align}
This output state consists of the sum of a purely uncorrelated part with a part that features correlations between the photons, however, these parts are not orthogonal. Therefore, it is not trivial to show analytically that only a linear phase response occurs in the limit $\sigma_p \to 0$. Instead, given that the input state is purely uncorrelated, we plot the input-output overlaps in Fig.~\ref{fig:fockscattering}
for one and two-photon Fock states as a function of $\sigma_\mathrm{p}$ assuming a Lorentzian pulse shape,
\begin{equation}
    \alpha(\omega) = \sqrt{\frac{2}{\pi}}\frac{\sqrt{\sigma_\mathrm{p}^3}}{\sigma_\mathrm{p}^2 + (\omega - \omega_0)^2},
\end{equation}
for convenience, yet without loss of generality in the limit $\sigma_\mathrm{p} \to 0$, and resonance such that $\omega_A = \omega_0$. We find that these overlaps are purely real for all $\sigma_\mathrm{p}$, and are able to clearly identify the desired linear phase response,
\begin{equation}
    \left|\mathrm{out}\right\rangle_n = e^{i\arg\left\{t(\omega_0)\right\}}\left|\mathrm{in}\right\rangle_n,
\end{equation}
for $\sigma_\mathrm{p} \leq 0.001\Gamma$, where $\arg\left\{t(\omega_0)\right\} = \pi$ on resonance. Specifically, the magnitudes of the input-output overlaps are both $>0.99$ if this constraint is met.

% Bibliography
\bibliography{references}